\documentclass[twocolumn]{aastex631}
\usepackage{mathtools}

\shorttitle{GD\,1400}
\shortauthors{Amaro et al.}

\begin{document}

\title{Phase-resolved Hubble Space Telescope WFC3 Spectroscopy of Weakly-Irradiated Brown Dwarf GD\,1400 and Energy Redistribution--Irradiation Trends in Six WD–BD Binaries}

\correspondingauthor{Rachael Amaro}
\email{rcamaro@arizona.edu}

\author[0000-0002-1546-9763]{Rachael C. Amaro}
\altaffiliation{National Science Foundation Graduate Research Fellow}
\affiliation{Department of Astronomy and Steward Observatory, The University of Arizona, 933 North Cherry Avenue, Tucson, AZ 85721, USA}

\author[0000-0003-3714-5855]{D\'aniel Apai}
\affiliation{Lunar and Planetary Laboratory, The University of Arizona, 1629 E. University Blvd, Tucson, AZ 85721, USA}
\affiliation{Department of Astronomy and Steward Observatory, The University of Arizona, 933 North Cherry Avenue, Tucson, AZ 85721, USA}

\author[0000-0003-2969-6040]{Yifan Zhou}
\affiliation{Department of Astronomy, 530 McCormick Rd, Charlottesville, VA 22904, USA}

\author[0000-0003-3667-8633]{Joshua D. Lothringer}
\affiliation{3700 San Martin Drive, Baltimore, MD 21218}

\author[0000-0003-2478-0120]{Sarah L. Casewell}
\altaffiliation{STFC Ernest Rutherford Fellow}
\affiliation{School of Physics and Astronomy, University of Leicester, Leicester LE1 7RH, UK}

\author[0000-0003-2278-6932]{Xianyu Tan}
\affiliation{Tsung-Dao Lee Institute \& School of Physics and Astronomy, Shanghai Jiao Tong University, Shanghai 201210, People’s Republic of China}

\author[0000-0003-1487-6452]{Ben W. P. Lew}
\affiliation{Bay Area Environmental Research Institute and NASA Ames Research Center, Moffett Field, CA 94035, USA}

\author[0000-0002-7129-3002]{Travis Barman}
\affiliation{Lunar and Planetary Laboratory, The University of Arizona, 1629 E. University Blvd, Tucson, AZ 85721, USA}

\author[0000-0002-5251-2943]{Mark S. Marley}
\affiliation{Lunar and Planetary Laboratory, The University of Arizona, 1629 E. University Blvd, Tucson, AZ 85721, USA}

\author[0000-0002-4321-4581]{L. C. Mayorga}
\affiliation{The Johns Hopkins University Applied Physics Laboratory, 11100 Johns Hopkins Rd, Laurel, MD, 20723, USA}

\author[0000-0001-9521-6258]{Vivien Parmentier}
\affiliation{Laboratoire Lagrange, Observatoire de la C\^ote d’Azur, Universit\'e C\^ote d’Azur, Nice, France}

\begin{abstract}
Irradiated brown dwarfs offer a unique opportunity to bridge the gap between stellar and planetary atmospheres. We present high-quality \textit{HST}/WFC3/G141 phase-resolved spectra of the white dwarf + brown dwarf binary GD\,1400, covering more than one full rotation of the brown dwarf. Accounting for brightness variations caused by ZZ Ceti pulsations, we revealed weak ($\sim$1\%) phase curve amplitude modulations originating from the brown dwarf. Sub-band light curve exploration in various bands showed no significant wavelength dependence on amplitude or phase shift. Extracted day- and night-side spectra indicated chemically similar hemispheres, with slightly higher day-side temperatures, suggesting efficient heat redistribution or dominance of radiative escape over atmospheric circulation. A simple radiative and energy redistribution model reproduced observed temperatures well. Cloud-inclusive models fit the day and night spectra better than cloudless models, indicating global cloud coverage. We also begin qualitatively exploring atmospheric trends across six irradiated brown dwarfs, from the now complete ``Dancing with Dwarfs'' WD--BD sample. The trend we find in the day-side/night-side temperature and irradiation levels is consistent with efficient heat redistribution for irradiation levels less than $\sim$10$^9$ ergs/s/cm$^2$ and decreasing efficiency above that level. 

\end{abstract}

\keywords{}

\section{Introduction} \label{sec:intro}

\begin{deluxetable*}{llcl}
\tablecaption{Properties of the GD\,1400 binary system \label{tab:keyprops}}
\tablewidth{10pt}
\tablehead{
\colhead{Parameter} & \colhead{Units} & \colhead{Value} & \colhead{Reference}
}
\startdata
$a$ & Orbital Separation [R$_{\odot}$] & 1.935 & \citet{Casewell24b} \\
$D$ & Distance [pc] & 46.25 $\pm $0.07 & \citet{FarihiChristopher2004} \\
$P$ & Orbital Period [hr] & 9.97968 $\pm$ 0.00192 & \citet{Burleigh11} \\
$T_{\rm{eff,WD}}$ & Effective Temperature [K] & 11,939 $\pm$ 177  & \citet{Casewell24b} \\
log$_{10}$($g$) (WD) & Surface Gravity [cm s$^{-2}$] & 8.123 $\pm$ 0.046 & \citet{Casewell24b} \\
$M_{\rm{WD}}$ & Mass [M$_{\odot}$] & 0.680 $\pm$ 0.029 & \citet{Casewell24b} \\
$M_{\rm{BD}}$ & Mass [M$_{\odot}$] & 0.074 $\pm$ 0.007 & \citet{Casewell24b} \\
$R_{\rm{BD}}$ & Radius [R$_{\odot}$] & 0.099 $\pm$ 0.018 & This work \\
Cooling Age (WD) & [Gyr] & 0.46$^{+0.04}_{-0.03}$ & \citet{Casewell24b} \\
System Age    & [Gyr] & 1.76$^{+1.20}_{-0.56}$ & \citet{Casewell24b} \\ 
$R_{\rm{WD}}$ & Radius [R$_{\odot}$] & 0.0119$\pm$0.0007 & This work \\
$R_{\rm{BD}}$ & Radius [R$_{\odot}$] & 0.099 $\pm$ 0.0185 & This work \\
              & Radius [R$_{\rm{Jup}}$] & 0.985 $\pm$ 0.1846 & This work \\
$i$ & Inclination [degrees] & 59.2$^{+6.7}_{-1.3}$ & This work
\enddata
\end{deluxetable*}

The study of ultracool ($<$3000~K) atmospheres is a rapidly evolving field, offering exciting discoveries and motivating advancements in observing and modeling techniques. Among the most notable subjects in this area are hot Jupiters -- gas giants in astonishingly close orbits to their host stars. These exoplanets have garnered significant interest, in part due to their complex atmospheric trends and dynamics.

Specifically, hot Jupiters with equilibrium temperatures above $\sim$2000~K often exhibit less efficient heat redistribution, leading to significant temperature contrasts between their day and night sides \citep{Showman13, Zhang18}. In contrast, cooler hot Jupiters redistribute heat more effectively, resulting in smaller temperature differences between their day and night sides \citep{Perez-Becker13}.
Phase curve observations and general circulation models suggest hot Jupiters have strong eastward jet streams at their equators \citep{ShowmanGuillot02, ShowmanPolvani11}. Additionally, ultra-hot Jupiters ($T_{\rm{eq}}>2000$~K)  exhibit temperature inversions on their day sides \citep{Fortney08}, attributed to high-altitude absorbers like titanium oxide (TiO) and vanadium oxide (VO) \citep{Hubeny03}.

Despite these discoveries, several key questions surrounding hot Jupiter atmospheres remain, including the formation and evolution of condensate clouds, the efficiency of day-to-night heat redistribution, the extent and impact of thermal dissociation/recombination on the day-side, and the role of magnetism \citep{RogersKomacek14}. 
Several studies have investigated these questions as a function of fundamental properties including incident irradiation strength, rotation rate, drag timescale, surface gravity, and chemical composition \citep[]{TanShowman20_rotationWDBDs, Roman21, Beltz22, Komacek22, Zhang23, Tan24, Roth24}. However, fully characterizing trends in hot Jupiter atmospheres remains challenging. Since most hot Jupiters are found around main sequence stars \citep{Wang15}, they consequently have relatively low planet-star flux contrasts, making it more time consuming to achieve high signal-to-noise ratio (S/N) observations.

A proxy for understanding these ultracool atmospheres are short-period ($\lesssim$10 hours) white dwarf-brown dwarf (WD--BD) binaries. The irradiated brown dwarfs in these systems share important similarities to hot Jupiters --- they experience similar incident irradiation levels, share comparable chemical compositions, are both tidally-locked, and are governed by analogous atmospheric processes \citep{Burrows01, Showman13, Bailey14, Marley15, Zhang_2020, Showman20}. A key advantage of studying irradiated brown dwarfs as proxies for hot Jupiter atmospheres is their greater observational accessibility. Brown dwarf emission primarily peaks in the near-infrared (NIR), while white dwarf emission peaks in the ultraviolet (UV), resulting in a more favorable flux contrast in the NIR compared to hot Jupiter systems. Additionally, white dwarfs are generally featureless and non-variable, making it easier to isolate the unresolved brown dwarf companions from their host stars.

Additionally, the shorter orbital periods of these systems further enhance their observational utility, requiring fewer observations to capture a full orbit and, consequently, a full rotation of the irradiated brown dwarfs. While there are distinctions in surface gravities and potential differences in bulk compositions between irradiated brown dwarfs and hot Jupiters, these factors are secondary to the primary similarities that make irradiated brown dwarfs excellent proxies for studying hot Jupiter atmospheres.

\begin{figure*}
    \begin{center}
    \includegraphics[width=0.78\textwidth]{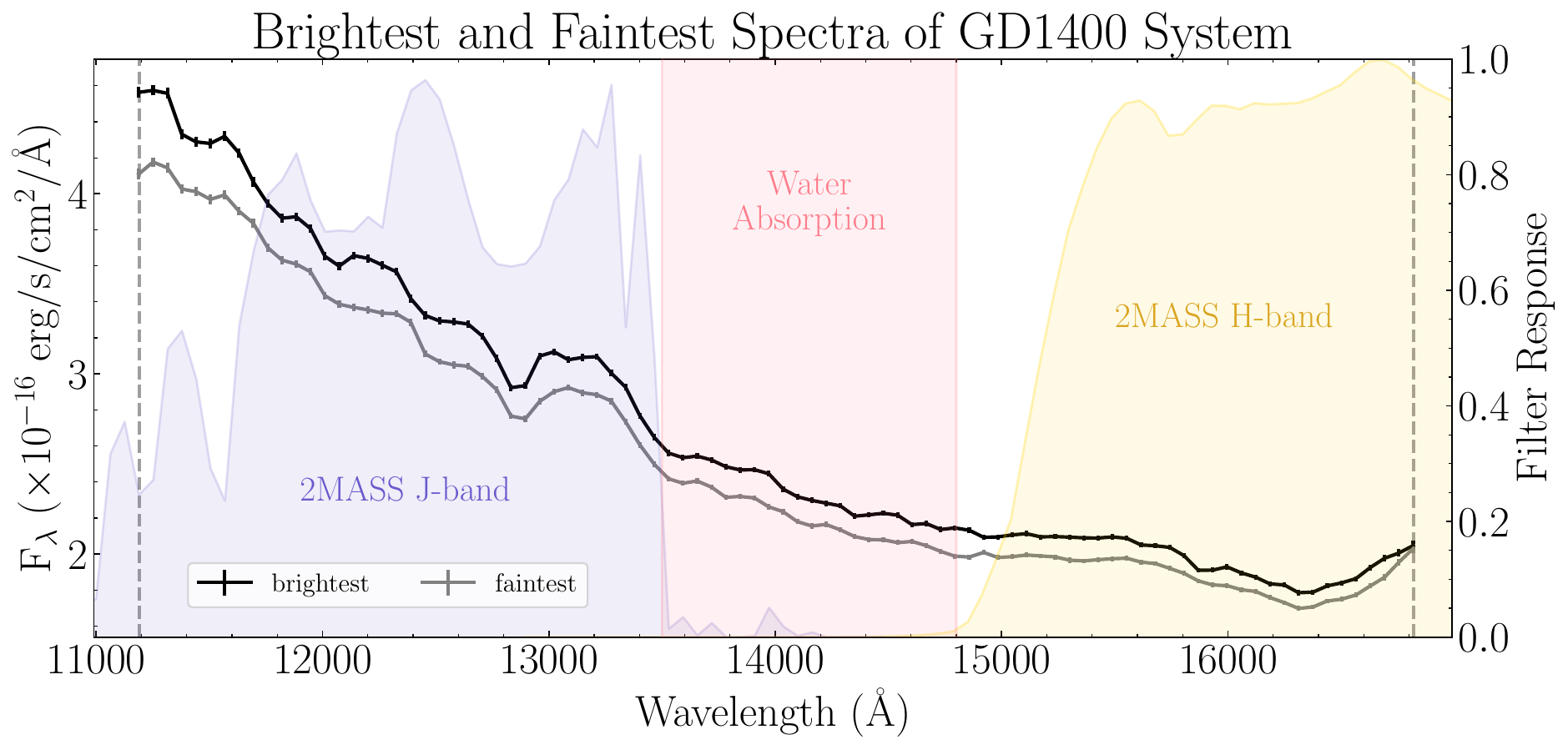}
    \caption{Brightest and faintest spectra of GD\,1400 from our \textit{HST} observations, after data reduction steps described in Section~\ref{sec:obs_dataredux}. Filter response profiles, used for sub-band light curve analyses, are shown as shaded regions with their corresponding labels.}
    \label{fig:allspec}
    \end{center}
\end{figure*}

However, some white dwarfs are not entirely non-variable due to intrinsic pulsations, which must be accounted for in detailed analyses. ZZ Ceti stars, also known as DAV white dwarfs, are pulsating white dwarfs with hydrogen-dominated atmospheres. These stars exhibit non-radial g-mode pulsations with periods ranging from 100 to 1500 seconds, driven by partial ionization of hydrogen in their outer layers \citep{Fontaine82, Winget88}. For WD--BD binaries, the presence of ZZ Ceti pulsations introduces additional variability to the system's light curve, potentially complicating the interpretation of brown dwarf phase-resolved observations. Careful isolation of these pulsations allows their effects to be mostly accounted for, allowing for analysis of the irradiated BD's atmosphere.

Short-period WD--BD binaries are a rare outcome of binary star evolution, with only a handful of known systems: GD\,1400 \citep{FarihiChristopher04, Dobbie05, Burleigh11}, WD0137-349 \citep{Maxted06, Burleigh06}, WD0837+185 \citep{Casewell12}, NLTT5306 \citep{Steele13}, SDSS J155720.77+091624.6 \citep{Farihi17}, WD1202-0242 \citep{Parsons17, Rappaport17}, SDSS J1231+0041 \citep{Parsons17}, EPIC212235321 \citep{Casewell18a}, SDSS1411+2009 \citep{Beuermann13, Littlefair14}, WD1032+011 \citep{Casewell20_WD1032}, Gaia J0052+4505 \citep{Owens23}, Gaia J0603+4518 \citep{Owens23}, and SDSS J222030.68-004107.3 \citep{Steele09, Casewell24b}. Fortunately, the irradiated brown dwarfs in these systems exhibit an irradiation temperature range ($T_{\rm{irr}} = 650 - 3400$~K) comparable to that of warm to ultra-hot Jupiters, facilitating the study of irradiated atmospheres as a function of incident irradiation levels.

One program working to address the key challenges in studying irradiated atmospheres is the Hubble Space Telescope (\textit{HST}) program GO-15947 (``Dancing with the Dwarfs'', PI: Apai), that has successfully observed six WD--BD binaries with phase-resolved spectroscopy from Wide Field Camera 3 (WFC3). Five irradiated BDs from this program have already been published. The first was SDSS1411B, which exhibited stronger water (1.4 $\mu$m) absorption on the night-side hemisphere and no evidence of a eastward-shifted hotspot \citep{Lew22}. In another study, \citet{Zhou22}, analyzed the phase-resolved spectra of WD0137B and EPIC2122B, two highly irradiated brown dwarfs. They found strong temperature contrasts between the day- and night-side hemispheres, with the day-side spectra being best fit to irradiated atmosphere models and the night-side spectra closely matching non-irradiated models (EPIC2122's night-side spectrum was only marginally detected and best-fit by a 1300~K blackbody). \citet{Amaro23} presented the results of NLTT5306B, which exhibited a small ($<100$~K) brightness temperature difference between the day and night hemispheres and was best fit by cloudy models for both hemispheres. Recently, results for SDSS\,1577B, a highly irradiated BD, were presented in \citet{Amaro24}. They showed evidence for inefficient day-to-night heat redistribution and a temperature inversion on the day-side.

In this work, we present phase-resolved \textit{HST} WFC3 observations of the WD--BD binary system GD\,1400, the sixth and final object in the ``Dancing with the Dwarfs'' program. GD\,1400 is the least strongly irradiated brown dwarf in the sample, with irradiation temperatures comparable to warm Jupiters. We also begin a comparison of the full Dancing with the Dwarfs WD--BD sample, exploring preliminary trends regarding phase-resolved brightness temperatures and incident irradiation strength.

\begin{deluxetable}{cllc}
\tablecaption{Log of HST WFC3 Observations for GD\,1400 \label{tab:obslog}}
\tablewidth{0pt}
\tablehead{
\colhead{Orbit} & \colhead{Observation} & \colhead{Filter} & \colhead{Exp. Start} \\
\colhead{} & \colhead{Type} & \colhead{ID} & \colhead{[BJD$_{\rm{TDB}}$ - 240000.0]}
}
\startdata
1 & Imaging & F132N & 59830.68513 \\
  & Spectroscopic & G141 & 59830.68625 \\
2 & Imaging & F132N & 59830.75136 \\
  & Spectroscopic & G141 & 59830.75248 \\
3 & Imaging & F132N & 59830.81749 \\
  & Spectroscopic & G141 & 59830.81862 \\
4 & Imaging & F132N & 59830.88363 \\
  & Spectroscopic & G141 & 59830.88475 \\
5 & Imaging & F132N & 59830.94979 \\
  & Spectroscopic & G141 & 59830.95092 \\
6 & Imaging & F132N & 59831.01592 \\
  & Spectroscopic & G141 & 59831.01704 \\
7 & Imaging & F132N & 59831.08205 \\
  & Spectroscopic & G141 & 59831.08318 \\
8 & Imaging & F132N & 59831.14820 \\
  & Spectroscopic & G141 & 59831.17308 \\
\enddata
\tablecomments{In each orbit, there were a total of two F132N direct images (22.3 seconds each) and 25 G141 grism observations (89.66 seconds each). BJD$_{\rm{TDB}}$ was converted from MJD using an online applet developed by Jason Eastman \citep[]{Eastman10_UTC2BJD}.}
\end{deluxetable}

First, in Section~\ref{sec:bdwd_system}, we provide a review on previous studies of GD\,1400. 
Next, in Section~\ref{sec:obs_dataredux} we present the HST time-resolved observations and data reduction methods.
We then present the results on the light curve analyses, including model fits, in Section~\ref{sec:lcanalysis}. 
In Section~\ref{sec:bdspec}, we analyze the rotational phase-resolved spectra, including the subtraction of the WD spectrum, extracting spectra as a function of rotational phase, and deriving day and night brightness temperatures. 
Then, in Section~\ref{sec:tempmap}, we present predictions on the day-to-night heat redistribution efficiency from a simple heat redistribution model.
In Section~\ref{sec:Compare_irr_models}, we compare the extracted BD spectra to one-dimensional radiative-convective irradiated atmosphere models.
Next, we begin a qualitative comparison of the full WD--BD HST sample in Section~\ref{sec:dwd_sample}, including all of the day- and night-side spectra and their derived brightness temperatures.
Finally, in Section~\ref{sec:conclusions}, we review the main conclusions of this study.

\section{White Dwarf Brown Dwarf Binary System GD 1400} \label{sec:bdwd_system}

There exist two independent spectroscopic analyses of the hydrogen-rich (DA) white dwarf GD 1400A. First, \citet{Koester01} analyzed high-resolution spectroscopy ($R=$18500) from the UV-Visual Echelle Spectrograph (UVES) on the Very Large Telescope (VLT) and derived $T_{\rm{eff}}=$ 11,604 $\pm$ 28~K, log($g$) $=$ 8.05 $\pm$ 0.02 for GD 1400A. Then, \citet{Fontaine03} observed GD 1400 with the high-resolution ($R\sim$ 805) EMMI instrument attached to the 3.6m New Technology Telescope (NTT) at ESO La Silla. They found $T_{\rm{eff}}=$ 11,550 $\pm$ 200~K, log($g$) $=$ 8.14 $\pm$ 0.05, in agreement with the values from \citet{Koester01}.

\citet{FarihiChristopher2004} presented evidence for GD\,1400B, the second known brown dwarf companion to a white dwarf, using $K$-band spectroscopy from NIRSPEC at Keck Observatory and optical $V$-band photometry from the Nickel 1~m telescope at Lick Observatory. The companion was discovered, then confirmed, through photometric excess and subsequent spectroscopy in the 2.2 $\mu$m region. Its apparent lack of excess emission at 1.2$\mu$m suggest a spectral type of L5.5 or later for the companion. Assessing its distance and accounting for NIR contributions of the WD, the absolute magnitude of GD 1400B would place it around spectral type L6 \citep{FarihiChristopher2004}. \citet{Dobbie05} estimated GD\,1400B to have a spectral type of L7 through simultaneous fits of the WD and BD components in an $HK$ grism observation, with model and empirical template spectra, respectively.

In a follow-up study, \citet{Farihi05} measured mid-IR fluxes for GD\,1400 using the Infrared Array Camera  (IRAC) on the \textit{Spitzer Space Telescope}. They calculated the expected flux from GD\,1400A from data in \citet{FarihiChristopher2004}, and subtracted this value to produce the contribution of GD\,1400B and further constrain the spectral type to be no earlier than L5. 

Radial velocity observations of GD\,1400 with UVES on European Southern Observatory’s Very Large Telescope (ESO VLT) conclusively show that this system is a post-common envelope binary with an orbital period $P$=9.98 hours \citep{Burleigh11}.

Recently, \citet{Casewell24b} obtained fifteen high resolution optical spectra of GD\,1400 from the UVES echelle spectrometer \citep{Dekker00_UVES} on UT2 of the ESO VLT and optical through mid-IR photometry from the South African Astronomical Observatory (SAAO), New Technology Telescope (NTT) Son OF ISAAC (SOFI), and Wide-field Infrared Survey Explorer ($WISE$). From the optical spectra, they measured radial velocities of the NLTE lines H$\alpha$ and H$\beta$ and conducted analysis on the time-series photometry, analyzing both the white dwarf pulsations and the effects of irradiation on the brown dwarf. The time-series photometry revealed that the observed pulsations are consistent with those of a large-amplitude ZZ Ceti variable, with frequencies and amplitudes that vary from year to year. Additionally, they conclude that GD\,1400B must have survived common envelope evolution during the WD's Asymptotic Giant Branch (AGB) phase in order to be in its present-day orbit, rather than the Red Giant Branch (RGB) phase like many of the other known WD--BD systems \citep[e.g. WD\,0137B;][]{Burleigh06}.

\section{Observations \& Data Reduction} \label{sec:obs_dataredux}

On 08 September 2022, we observed eight \textit{Hubble Space Telescope} (\textit{HST}) orbits of GD\,1400, under Cycle 27 program GO-15947 (PI: Apai). Each orbit consisted of 25 spectral observations using the G141 grism and two direct images using the F132N filter. A log of the \textit{HST} observations is presented in Table~\ref{tab:obslog}.

\begin{figure*}
    \begin{center}
    \includegraphics[width=0.99\textwidth]{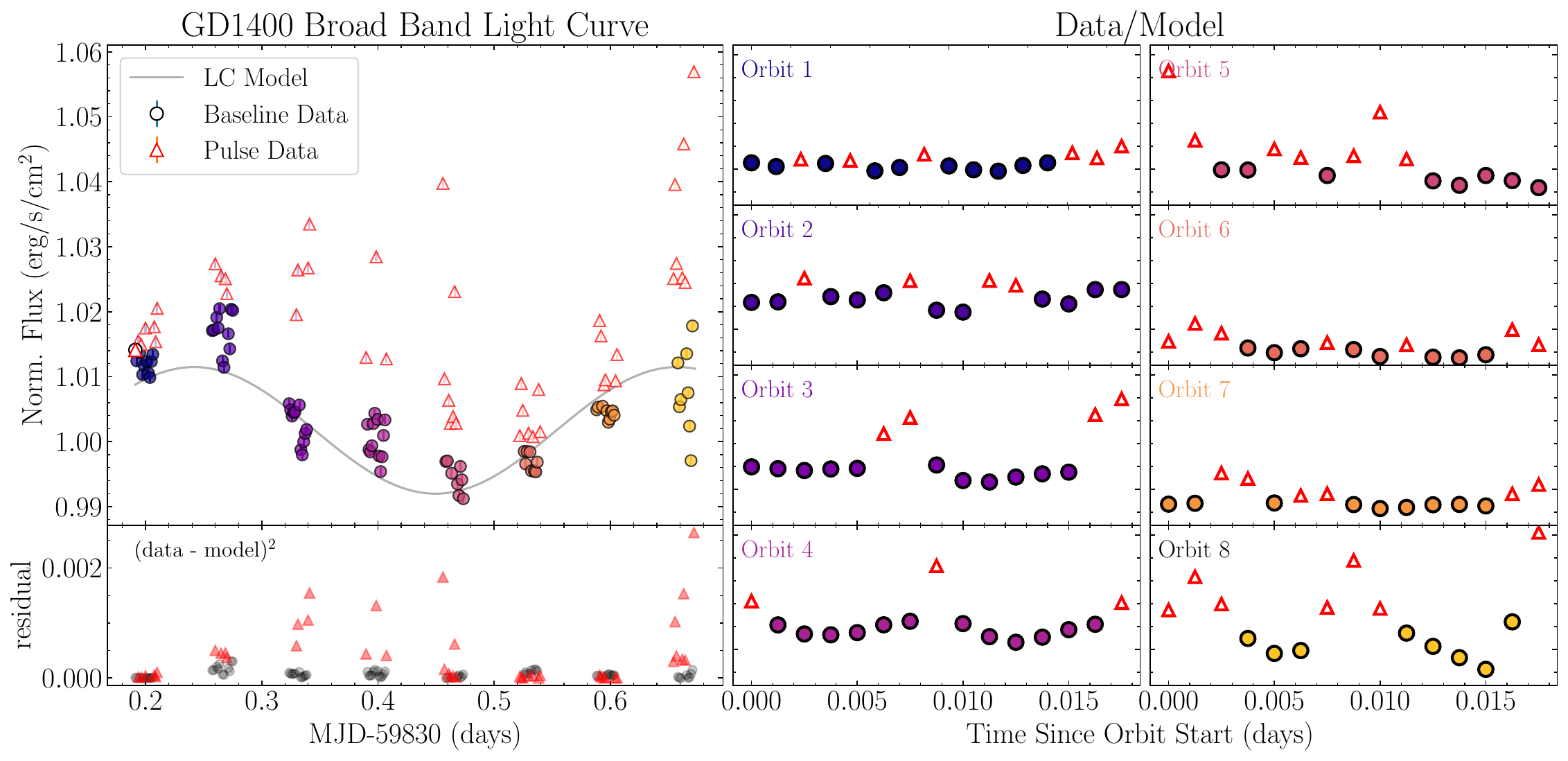}
    \caption{\textit{Left:} Broadband light curve of GD\,1400 with best-fit light curve model, normalized to the median value of the faintest third epochs in each orbit. This low-order model represents the flux modulations owing to the tidally-locked orbit of the companion BD. Epochs classified as ``baseline" are shown as black outlined markers, while epochs classified as ``pulse" are shown as empty, red triangles. Errorbars are present, but smaller than the marker sizes. \textit{Right:} GD\,1400 light curve with the BD model divided out, revealing 2--3 identified pulses per orbit.}
    \label{fig:bblc_pulses}
    \end{center}
\end{figure*}

At the beginning of each orbit, two direct images were obtained for wavelength calibration using the F132N filter, a 256$\times$256 subarray setup, and the GRISM256 aperture. After direct images, each orbit acquired twenty-five spectroscopic exposures in “Staring mode” of $\sim$90 s each, using the G141 grism, a 256×256 subarray setup, and the GRISM256 aperture. This observing sequence was successfully carried out for five previous WD--BD systems as part of the same HST program: WD\,0137, EPIC\,2122 \citep{Zhou22}, SDSS\,1411 \citep{Lew22}, NLTT5306 \citep{Amaro23}, and SDSS\,1557 \citep{Amaro24}.

The phase-resolved spectra were extracted using an established pipeline that combines the WFC3/IR spectroscopic software \texttt{aXe} \citep{Kummel09_aXe} and custom \texttt{Python} scripts. This pipeline has been successful at  extracting time-resolved observations of brown dwarfs \citep[e.g.][]{Buenzli12, Apai13} as well as time-resolved observations of WD--BD binaries \citep[e.g.][]{Zhou22, Lew22, Amaro23}. Those studies have comprehensively described the data reduction steps, but here we also provide a summary.

To start, we downloaded the \texttt{flt} files from the Barbara A. Mikulski Archive for Space Telescopes (MAST)\footnote{\url{https://archive.stsci.edu/index.html}}. We then organized the files into their respective orbits. For the G141 data, we corrected bad pixels, identified via the Data Quality (DQ) extension, by linearly interpolating with adjacent good pixels. The linear interpolation was performed in the x and y direction using the median value of the neighboring four pixels on each side (i.e., up to sixteen pixels). The F132N images and G141 data were then embedded into larger full-frame (1,014 $\times$ 1,014) arrays.

Next, we constructed a median combined direct image by taking the median of each pixel across all full-frame F132N direct images and performed precise source extraction \citep{Bertin96_sextractor}. The output catalog was used by \texttt{aXe} for wavelength calibration. Next, we ran \texttt{axeprep} on the G141 spectroscopic images to perform background subtraction, exposure time normalization, and gain correction. Interestingly, we noticed that the last 5-7 exposures in each orbit had increased background levels, becoming brighter with each consecutive exposure. With the assistance of the HST Help Desk\footnote{\url{https://stsci.service-now.com/hst}}, we determined that this was likely caused by the illuminated Earth limb coming within 50 degrees to the line-of-sight of \textit{HST} on 08 September 2022. To eliminate any Earth contamination, we discarded the affected G141 files, leaving 121 observations suitable for analysis. 

\begin{figure}
    \begin{center}
    \includegraphics[width=0.48\textwidth]{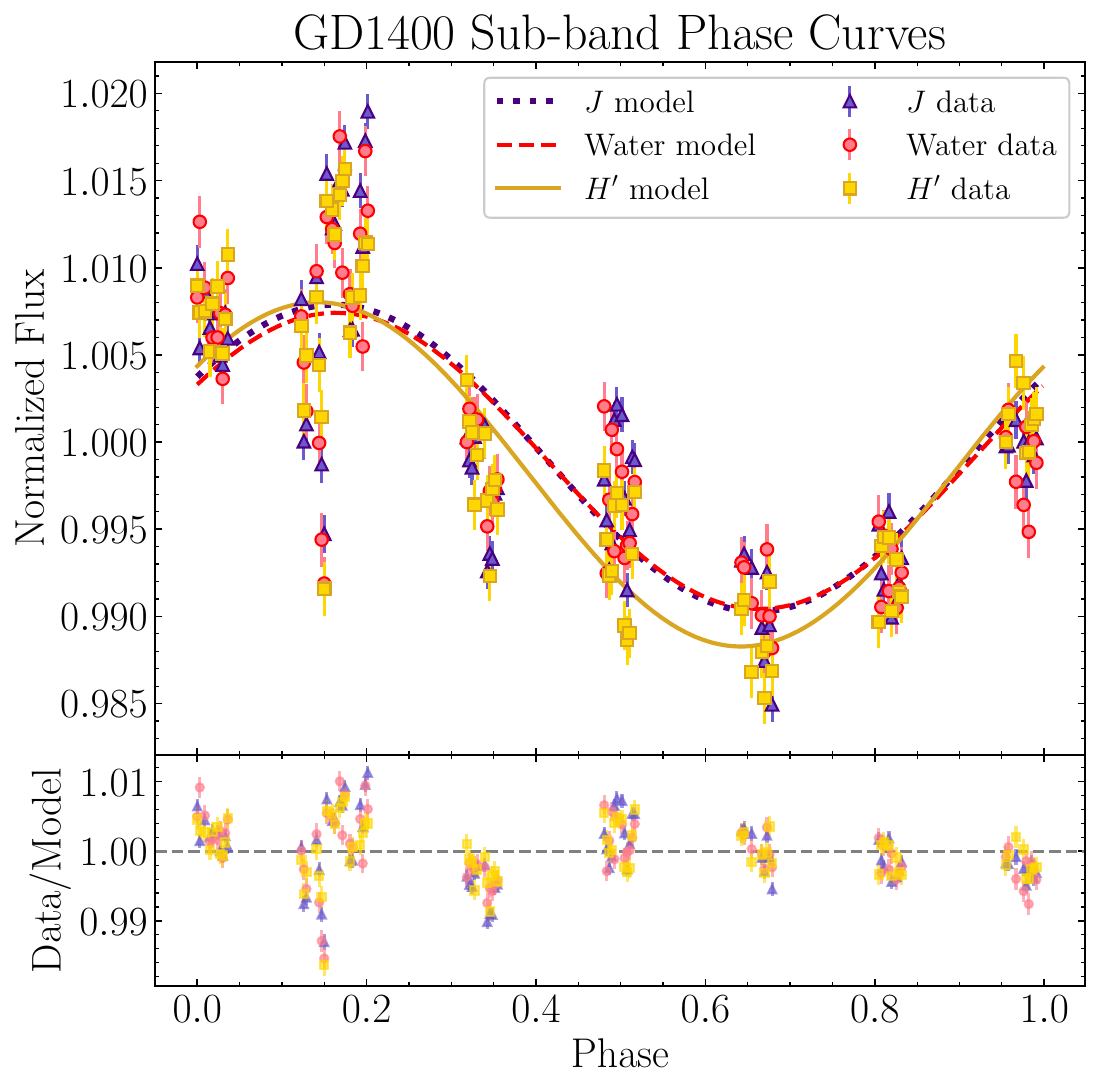}
    \caption{Phase-folded sub-band light curves in the $J$ (indigo triangle), Water (pink circle), and $H'$ (golden square) bands along with best-fit phase curves as dashed, dotted, and solid lines, respectively. A phase of 0 corresponds to the first HST observation. We attempted higher order phase curve fits, but decided that the unstable WD pulsations might contaminate the fits. We observed a higher amplitude in the $H'$-band phase curve relative to the $J$ and Water bands.}
    \label{fig:subband_phasecurve}
    \end{center}
\end{figure}

After \texttt{axeprep}, the spectra were extracted with \texttt{axecore} and corrected for ramp effect using \texttt{RECTE} \citep{Zhou17}. Finally, we resampled the spectra to the actual resolving power of our observations by modeling the spectral response function from each 2D stamp output by \texttt{hstaxe}. The spectral response function was measured by plotting the counts of each pixel in the perpendicular direction to the dispersion axis and fitting the shape of this curve with Moffat profile. The Full-Width-Half-Maximum (FWHM) of this Moffat profile was determined to be the actual resolving power. For GD\,1400, the resulting resolution of the spectra was R$\sim$132 at 1.4 $\mu$m, in agreement with the expected resolving power for the G141 grism\footnote{\url{https://hst-docs.stsci.edu/wfc3ihb/chapter-8-slitless-spectroscopy-with-wfc3/8-1-grism-overview}}. Examples of the final science spectra, shown with the brightest and faintest spectra among our observations, are presented in Figure~\ref{fig:allspec}.

\section{Light Curves} \label{sec:lcanalysis}
The tidally-locked orbits of irradiated brown dwarfs and many hot Jupiters implies that observing these systems at various orbital phases will directly map to the rotational phase of the companion. For most WD--BD systems, we assume constant flux from the white dwarf primary, meaning that any periodic changes in the observed flux likely comes from longitudinal intensity modulations present in the photosphere of the irradiated brown dwarf.

However, It is important to note that GD\,1400A is a known ZZ Ceti pulsator with multiple published periods: 823.2, 727.9, and 462.2 s \citep{Fontaine03}; 730 and 454 s \citep{Kilkenny14}; and 1046, 796, and 766 s \citep{Bognar20}. Given the red-end position of GD\,1400A within the instability strip, we do not expect the frequencies of these pulsations to be stable over time, making exact predictions for pulse timing a near impossible task. Thankfully, the measured pulsations are much shorter than the orbital period of GD\,1400B, meaning the WD pulsations should not affect the best-fit period of our light curve models, further detailed in this section.

\begin{figure}
    \begin{center}
    \includegraphics[width=0.4\textwidth]{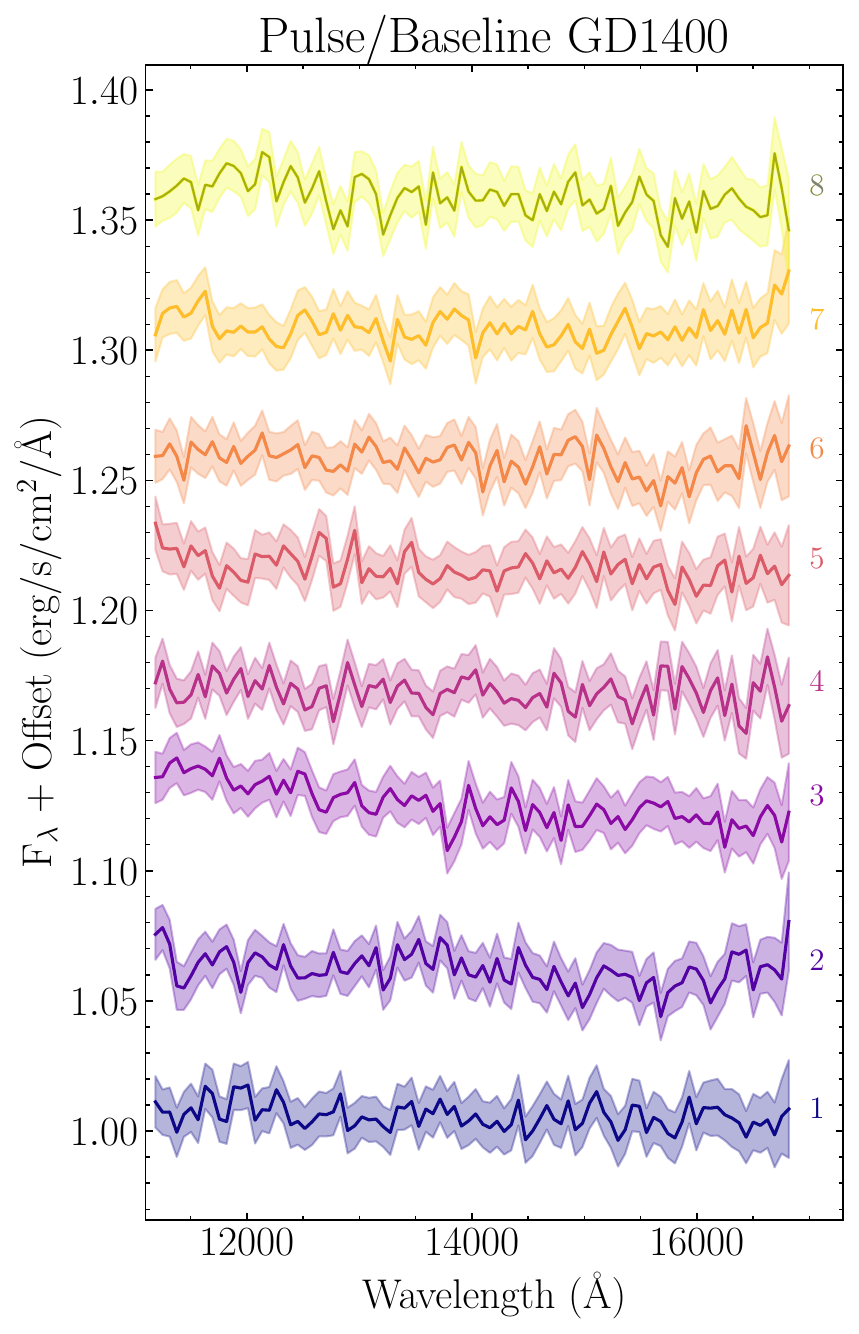}
    \caption{Flux ratios in each HST orbit (labeled on the right) for epochs identified as ``in pulse'' vs. baseline. Shaded regions depict propagated uncertainty and ratios are offset for visual clarity. Only the flux ratio in Orbit 3 exhibited a slight slope, but within 1-$\sigma$ of the other orbits, and we didn't observe any systematic features at consistent wavelengths. These findings suggest that the ``in pulse'' epochs are more of a brightness scaling rather than a change in relevant chemical abundances.}
    \label{fig:pulse_spec}
    \end{center}
\end{figure}

\subsection{Broad-band Analysis} \label{sec:bb_lc}

To characterize the flux modulations coming the rotation of GD\,1400B, we created a simple light curve model designed to capture most of the periodic behavior. Following Equation~1 in \citet{Amaro24}, the model is a combination of Fourier series with four free parameters: scaling for normalization, orbital period, and amplitudes for the sine and cosine terms.

Additionally, to minimize any influence of WD pulsations on the best-fit light curve amplitudes intended to measure only the BD's rotation, we filtered out data points likely to be affected by pulsations. We achieved by classifying each light curve point as either “in pulse” or “baseline,” based on its flux level relative to others within the same orbit. Specifically, we assumed that the lowest flux levels within each orbit were least affected by pulsations and established the following method to identify “in pulse” epochs: (1) we calculated the median of the faintest third of light curve point in each orbit, to approximate a baseline flux and avoid the impact of flux outliers; (2) we measured the standard deviation ($\sigma_{\rm{STD}}$) across all epochs; and (3) any light curve points above \texttt{median}$+\frac{1}{3}\sigma_{\rm{STD}}$ were classified as “in pulse.” This threshold allowed us to systematically filter out flux levels most likely affected by pulsations, leaving a conservative estimate of the “baseline” flux.

This method led to 44 of the 121 epochs (36\%) being classified as “in pulse,” as shown in Figure~\ref{fig:bblc_pulses}. Importantly, even with these points excluded, the remaining “baseline” epochs provided full phase coverage, enabling us to fit a reliable light curve model for the brown dwarf’s rotation using an MCMC approach, also shown in Figure~\ref{fig:bblc_pulses}. Although this approach identifies “in pulse” points in all orbits, the chosen threshold mitigates false positives by excluding only those epochs with significant flux increases. Furthermore, minor misclassifications of baseline points as “in pulse” would not significantly alter our phase coverage, ensuring robust conclusions on the brown dwarf’s rotational modulation.

\begin{figure*}
    \begin{center}
    \includegraphics[width=0.7\textwidth]{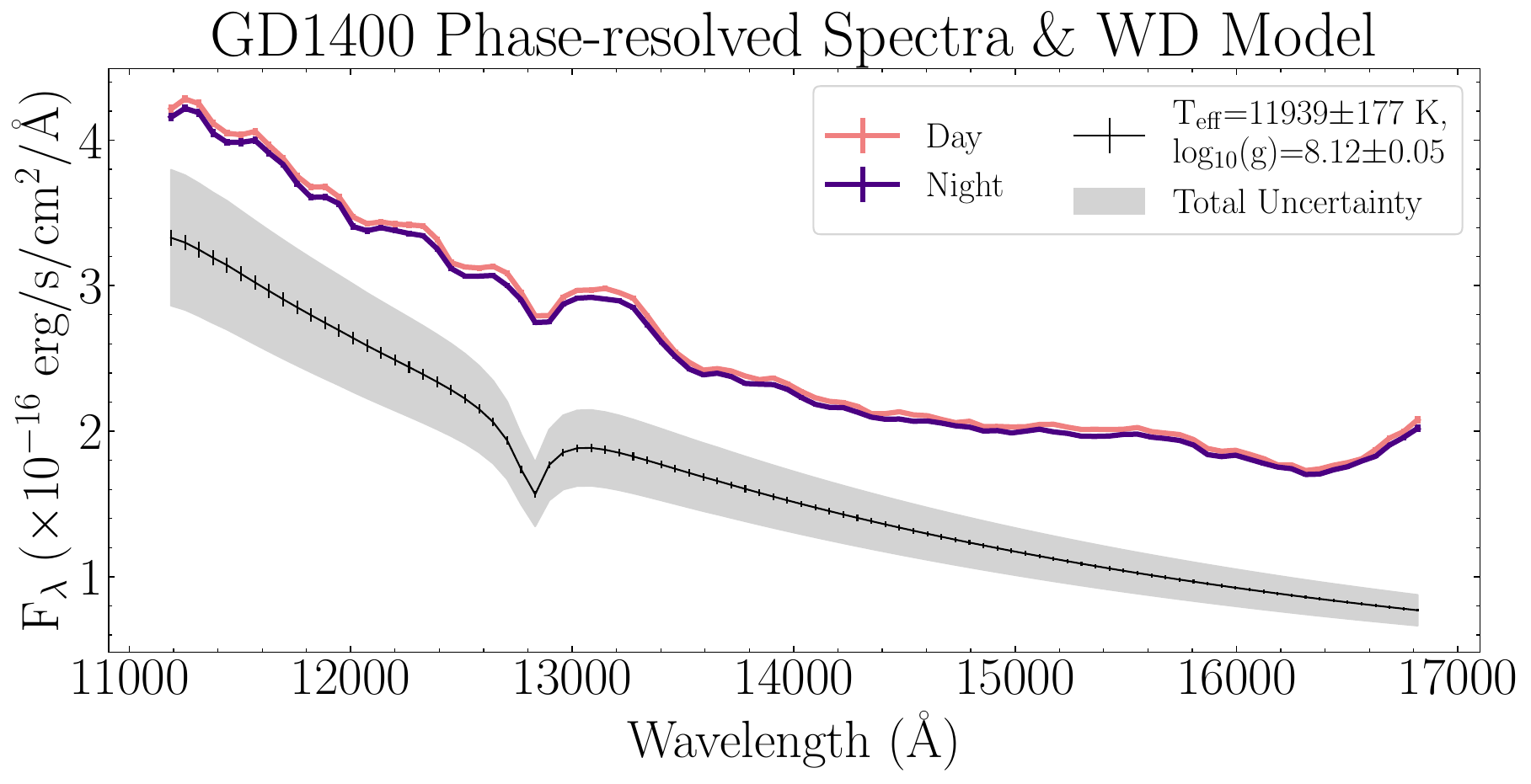}
    \caption{Modeled flux contribution of the primary WD GD\,1400A, shown with two sources of uncertainty: the first being from $T_{\rm{eff}}$ and log$_{10}$($g$) during the creation of the model (black error bars) and the second being from $R_{\rm{BD}}$ and distance, which affects the flux scaling (gray shaded region). Additionally, derived day (pink) and night (purple) spectra of the irradiated BD GD\,1400B are shown with 1-$\sigma$ uncertainties.}
    \label{fig:bdspec_wdmodel}
    \end{center}
\end{figure*}

Given that the data contains a broad range of low- to high-order frequencies owing to both the BD orbit and the ZZ Ceti pulsations, we decided to constrain the BD broadband light curve to the lowest order combination of Fourier series, i.e., only allowing a $k$=1 fit. The resulting broadband light curve model, shown in the top left panel of Figure~\ref{fig:bblc_pulses}, yielded a period of 0.415744$\pm$0.000347 days (9.9779$\pm$0.0083 hr), which was in agreement with the value published by \citet{Burleigh11}. The total amplitude was calculated following Equation (2) in \citet{Zhou22},
\(amp_k = \sqrt{a_k^2+b_k^2}\),
where $a_k$ and $b_k$ were the amplitudes for the sine and cosine terms for each $k$ order, respectively. Following this equation, GD\,1400 exhibited an amplitude of 0.974$\pm$0.038 percent, the lowest measured amplitude in our program: EPIC2122 exhibited 29.1\% \citep{Zhou22}, SDSS\,1557 10.5\% \citep{Amaro24}, WD0137 5.27\% \citep{Zhou22}, SDSS\,1411 3.8\% \citep{Lew22}, and NLTT5306 6.0\% \citep{Amaro23}. However, this result could be expected, as GD\,1400B orbits its WD primary at a distance of approximately 0.0089 AU (Table~\ref{tab:keyprops}), which is the farthest separation in our sample, compared to the closest system, EPIC/,2122, with a separation of 0.002 AU.

\subsection{Sub-band Analysis} \label{sec:sb_lc}

In addition to broadband light curve analyses, we studied intensity changes as a function of phase using similar methods in other studies \citep[e.g.][]{Buenzli12, Apai13, Buenzli14}. Namely, we derived sub-band light curves by convolving our spectra with different transmission filters , shown as shaded regions in Figure~\ref{fig:allspec}. First, we multiplied the Two-Micron All Sky Survey (2MASS) $J$-band transmission filter from the 2MASS All-Sky Data Release \citep{2MASS_Skrutskie06} over our spectra from 11000 to 13500 \AA{}, creating a sub-band $J$ light curve.
Similarly, we did the same with the 2MASS $H$-band transmission filter, with a key difference being that the $H$-band filter was truncated at 16700 \AA{}, since this is where our HST spectra ended (see Figure~\ref{fig:allspec}).
Essentially, we created a modified $H'$-band light curve from $15000-16700$ \AA{}.
Finally, we investigated the water band centered around $1.4~\mu m$ in order to further understand the atmospheric structure and thermal properties of the GD\,1400B. As the most abundant molecule after hydrogen and helium in both substellar and planetary atmospheres, water is a key tracer offering insights into temperature distribution, cloud dynamics, and chemical composition \citep[e.g.][]{Cushing08, Line17, Kreidberg18, Manjavacas18}.
We created a water band light curve simply by integrating from 13500 to 14500 \AA{}, i.e. no transmission filter was applied. The resulting sub-band light curves, presented in Figure~\ref{fig:subband_phasecurve}, were then normalized by their median flux.

Different wavelengths probe different pressure regions \citep[e.g.,][]{Buenzli12, Yang16}. Thus, studying sub-band light curves allow us to identify pressure-dependent structure and dynamics within an atmosphere. To quantify any similarities or differences between our three sub-band light curves (i.e., $J$, water, and $H'$ band), we created phase-curve models using the same approach as described in Section~\ref{sec:bb_lc}. The resulting best-fit phase curve models for each sub-band are shown in Figure~\ref{fig:subband_phasecurve}. We observed that the $J$-- and Water--band phase curve amplitudes were similar, 1.761$\pm$0.035 and 1.699$\pm$0.049 percent, respectively, but the $H'$--band exhibited a larger amplitude of 1.975$\pm$0.049 percent, suggesting the $H'$--band probed a pressure region exhibiting higher variability.

\section{Rotational Phase-Resolved Spectra} \label{sec:bdspec}

\subsection{Influence of Pulse vs. Baseline states} \label{pulse}

Figure~\ref{fig:pulse_spec} shows the ratio of ``in-pulse'' spectra versus baseline spectra for each orbit. If the ``pulse'' state caused a significant spectral transformation, e.g. emission features only seen in the ``pulse'' state or a hotter, bluer, blackbody contribution from the WD, then this would be visible in the ratios. Inspecting Figure~\ref{fig:pulse_spec}, we found no systematic changes in spectral features and only Orbit 3 exhibited a weak slope. No changes in spectral features between the ``pulse'' and ``baseline'' spectra indicated no significant chemical transitions, at least not significant enough to affect our analysis. Additionally, the lack of a slope in nearly all Orbits suggested that the temperature did not significantly change between states. We interpreted this to mean that a ``pulse'' essentially only has a white light effect. Therefore, spectral corrections were not necessary when later extracting the BD signal from the phase-resolved WD--BD spectra. 

\subsection{Modeling the White Dwarf Contribution} \label{sec:wdmodel}
In order to characterize the irradiated brown dwarf atmosphere, we needed to isolate and extract its signal from the WD--BD spectra. Since this system is unresolved and does not eclipse, we can never observe the brown dwarf companion alone. Thus, we modeled and subtracted the WD contribution from all spectra using a model grid of pure-hydrogen atmosphere WDs from \citet{Koester10} that takes effective temperature, $T_{\rm{eff}}$, and surface gravity, log($g$), as free parameters. To capture uncertainty in $T_{\rm{eff}}$ and log($g$), we generated 10,000 model spectra and sampled the two free parameters from Gaussian distributions centered on 11,939~K and 8.123 cm~s$^{-2}$, with standard deviations of 117~K and 0.046 cm~s$^{-2}$, respectively. At each wavelength, the full width at half maximum (FWHM) of the fluxes from these models gives the uncertainty in the WD flux, shown as black error bars in Figure~\ref{fig:bdspec_wdmodel}. Additionally, to account for uncertainty in the scale factor required to convert model fluxes to observed fluxes, we scaled the model flux $\frac{R_{\rm{WD}}}{D}^2$, propagating errors for from the WD radius, $R_{\rm{WD}}$=0.012$\pm$0.001 R$_{\odot}$, and distance $D$=46.24$\pm$0.07 pc. The resulting total uncertainty is represented by the gray shaded region in Figure~\ref{fig:bdspec_wdmodel}.

\begin{figure}
    \begin{center}
    \includegraphics[width=0.47\textwidth]{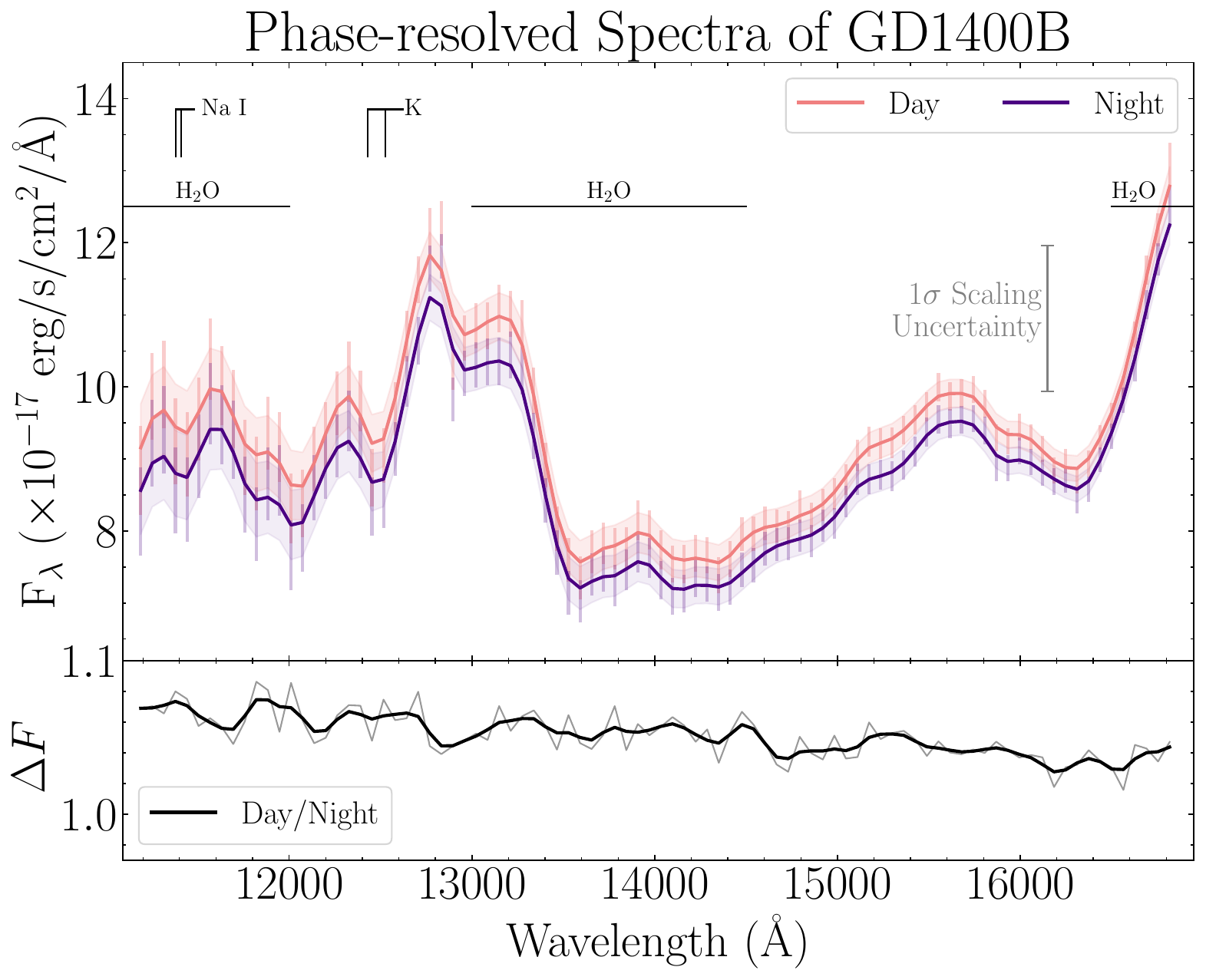}
    \caption{\textit{Top:} Extracted spectra of Brown Dwarf GD\,1400B, representing day (pink) and night (purple) hemispheres, where we expect to see the greatest variance in temperature, chemistry, etc. A gray, vertical representing $1\sigma$ scaling uncertainty is also shown, which would essentially move the spectra up or down together, and would not affect the ratio between the spectra. Wavelengths where we could expect water, K, and Na\,\textsc{I} absorption are labeled. We observed no significant differences in spectral features, suggesting that the two hemispheres are chemically similar. \textit{Bottom:} A ratio between the day and night spectra, revealing a weak, negative slope that could indicate a relatively bluer day-side spectrum, or conversely a redder night-side spectrum. Reasons for this could include hotter emission from the day-side and/or changes in the absorption or scattering features between the two hemispheres.}
    \label{fig:bdspec}
    \end{center}
\end{figure}

\begin{figure}
    \begin{center}
    \includegraphics[width=0.45\textwidth]{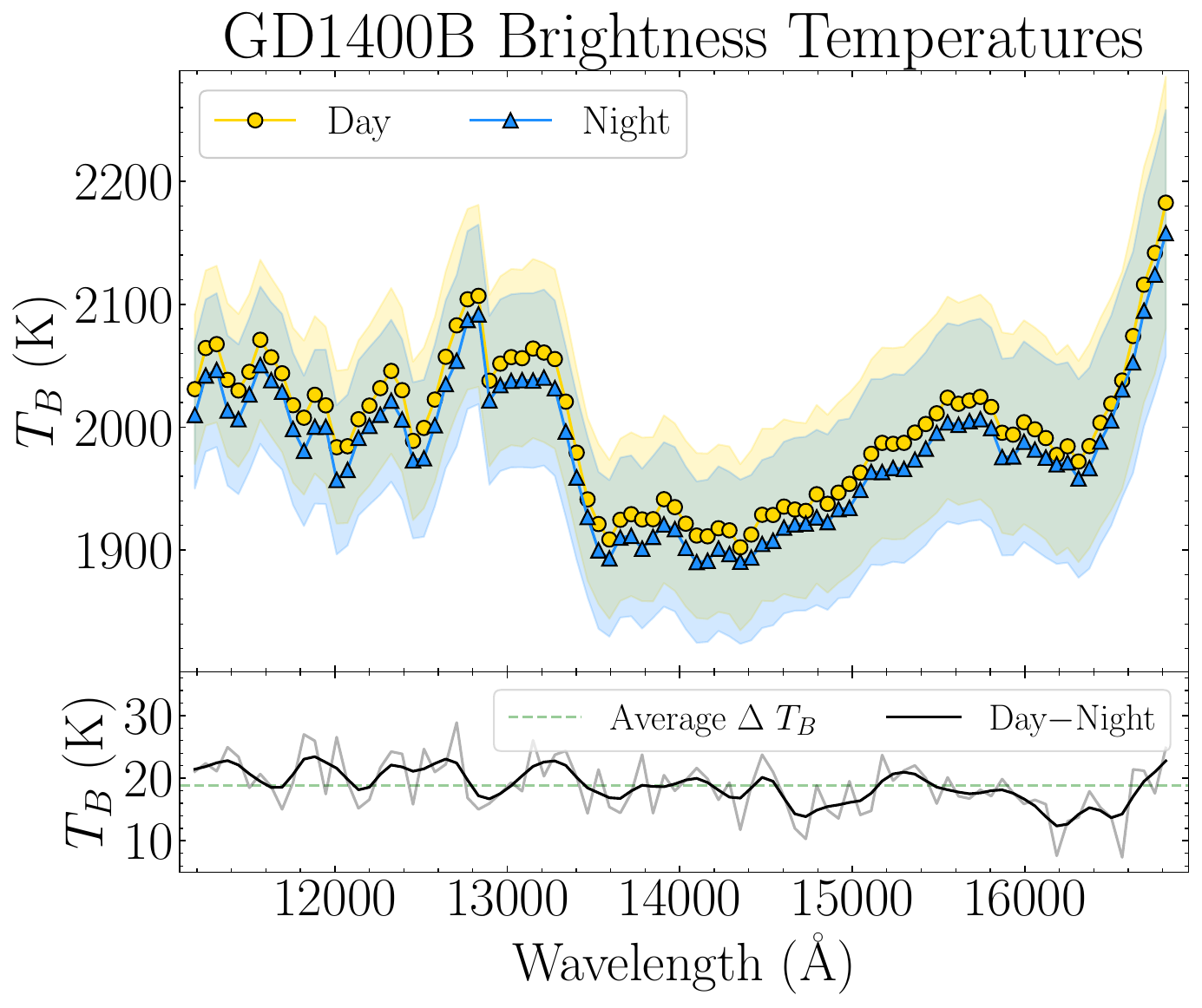}
    \caption{Brightness temperatures for the day and night sides of GD\,1400B, exhibiting nearly identical, strong wavelength dependence. Shaded regions behind each curve depict 1-$\sigma$ uncertainty. Average Day$-$Night temperature difference is 19$\pm$102~Kelvin, with small deviations across the wavelength range, suggesting statistically similar temperature structures on both day and night hemispheres.}
    \label{fig:brighttemps}
    \end{center}
\end{figure}

\subsection{Extracted Phase-Resolved Spectra of GD\,1400B} \label{bdspec}
To study the spectra of GD\,1400B as a function of phase, we started by isolating the two extreme hemispheres, defined as Day and Night, since differences should be the most intense at these phases. To ensure data quality and minimize spectral variations, we created Day and Night spectra with median combinations of five spectra each, selecting the five brightest spectra in the sample for Day and the five faintest spectra for Night. We also confirmed that the phases of the selected spectra were within $\pm$5 degrees of the light curve maximum and minimum, respectively, to maximize any possible hemisphere-dependent signatures. The resulting hemisphere-integrated Day and Night spectra for the WD--BD system can be seen in Figure~\ref{fig:bdspec_wdmodel}.

Next, we subtracted the WD model from the Day and Night spectra and isolated the BD contribution for further atmospheric investigation, as shown in top panel of Figure~\ref{fig:bdspec}. Both hemispheres exhibit broad water absorption around 1.4~$\mu$m, as well as Na\,\textsc{I} (1.138 and 1.141~$\mu$m) and K I (1.243 and 1.252~$\mu$m) absorption. Notably, these features do not significantly change between the day- and night-side hemispheres, suggesting efficient atmospheric circulation or insufficient irradiation to drive a thermal inversion \citep{Fortney08, Parmentier16}. Across our HST wavelengths of $\sim$1.1 to 1.67 $\mu$m, the Dayside ranges declined from 8 percent brighter at the bluest end to only 2.5 percent at the reddest end, with an average ratio of $\sim$5 percent. The greater difference at bluer wavelengths is perhaps expected, given the Dayside is irradiated by a much bluer (UV/Optical) WD primary. This slight slope in the Day/Night ratio suggests that the heat redistribution from Day-to-Night is not 100 percent. Constraining the redistribution efficiency is explored more in Section~\ref{sec:tempmap}.

\begin{figure}
    \begin{center}
    \includegraphics[width=0.48\textwidth]{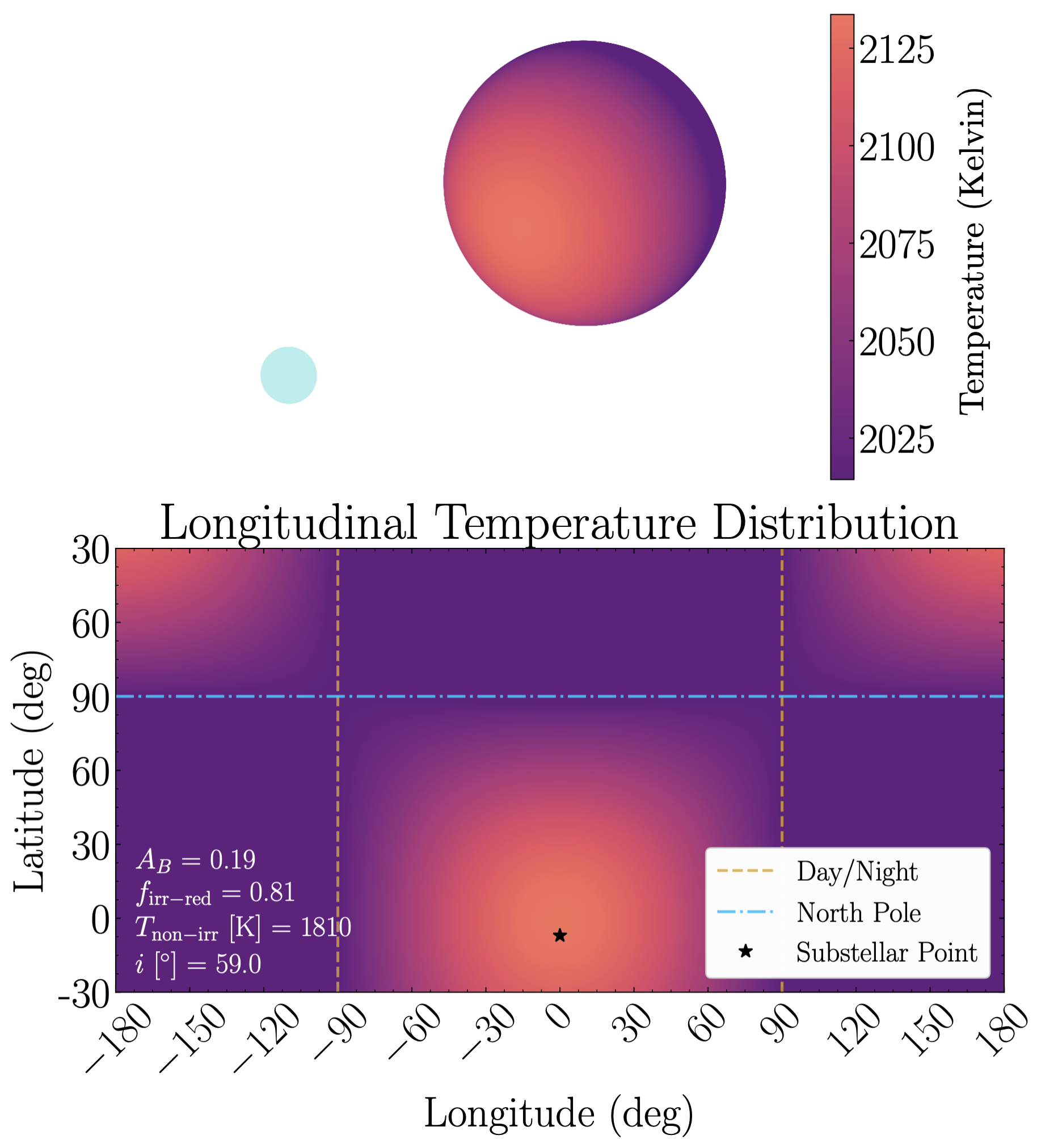}
    \caption{\textit{Top:} 3D projection of the temperature distribution on GD\,1400B, created by the simple heat redistribution model in Section~\ref{sec:tempmap}. Relative radii of the WD (pale blue circle) and BD are to scale, but the orbital separation is zoomed in. \textit{Bottom:} Full 2D projection of the temperature map, with best-fit values labeled in bottom left. Day/Night terminators (yellow dashed lines), the North pole (blue dot-dashed line), and the substellar point (black star) are shown as well, to visualize how the inclination affects our observations.}
    \label{fig:tempmap}
    \end{center}
\end{figure}

\subsection{Day and Night Brightness Temperatures of GD\,1400B} \label{brighttemps}
The combination of internal heat flux, net absorbed stellar flux, and opacity structure controls the thermal structure in the atmosphere of a stellar companion \citep{Marley15}. To study the thermal structure of GD\,1400B in detail, and thus better understand the influence of its atmospheric processes, we derived brightness temperatures as a function of wavelength. Following the Planck equation and solving for $T_{\rm{B}}$ at each wavelength, we calculated brightness temperatures for both the Day and Night hemispheres, shown in the top panel of Figure~\ref{fig:brighttemps}.

The brightness temperatures profiles for both hemispheres are strikingly similar, with nearly identical wavelength dependence, temperatures that range from 1880 to 2180, and an average difference of 19$\pm$102 Kelvin across the wavelength range. The statistical similarity between the two spectra suggests that the gas-phase abundances and opacity sources on GD\,1400B do not significantly vary as a function longitude, possibly due to strong heat redistribution from the WD primary's irradiation, internal heat that dominates the atmospheric dynamics over external irradiation, or some combination of both.

\section{Longitudinal Temperature Distribution} \label{sec:tempmap}
Given the complex nature of atmospheric processes---encompassing energy redistribution, reflectivity, and residual internal heat from formation---understanding the interplay and relative contributions of each is essential towards building a complete picture of a given atmosphere. Following \citet{Amaro23}, we created a simple energy redistribution model that simulates an irradiated, tidally-locked surface and redistributes the energy from day to night according to four atmospheric and orbital properties. These properties are: (1) Albedo, $A_B = [0.0, 1.0]$ in increments of $0.025$, (2) the irradiation redistribution fraction, $f_{\rm{irr-red}} = [0.0, 1.0]$ in increments of 0.02, (3) the non-irradiated brown dwarf temperature, $T_{\rm{non-irr}}$, sampled from 800 to 2000 K in increments of 25 K, and (4) the inclination, $i$, from 50 to 100 degrees in increments of 0.5. The inclination was allowed to go over 90 degrees for uncertainty calculations, but effectively, anything over 90 was wrapped to be within 0 and 90 according to $90 - (i_{90+} - 90)$. We conducted a grid search among all possible parameter combinations, and then created each model following a few simple steps.

First, we applied a uniform day- and night-side temperature determined by the input $T_{\rm{non-irr}}$. Then, the incident radiation from the WD was added to the day side, and the value of $A_B$ determined how much was reflected away. In this tidally-locked setup, the substellar point is the hottest location on the BD surface for all $A_B < 1.0$. Next, we redistributed a fraction of the incident radiation from the day-side, determined by $f_{\rm{irr-red}}$, and evenly distributed it onto the BD entire surface to preserve continuity at the day/night terminators. If $f_{\rm{irr-red}} = 1.0$, all irradiated flux would be evenly redistributed, resulting in a uniform temperature map. Finally, we added the effect of inclination by taking a number of rows at the top of the temperature map and switching them between day and night hemispheres. The number of rows were determined by the inclination parameter, $i$, where a lower $i$ would result in a larger amount of rows, due to the greater percentage of the opposite hemisphere being seen.

To find the best-fit model for GD\,1400B, we calculated residuals between the observed brightness temperatures and the hemisphere-integrated temperatures in each model. Given the observed wavelength dependence on the brightness temperatures (Figure~\ref{fig:brighttemps}), we used an inverse variance-weighted average temperature from 1.23 to 1.32 $\mu$m, based on the wavelength ranges used in \citet{Zhou22}. In this wavelength range, we targeted the continuum pressures below where the atmosphere would react strongly to irradiation, e.g. temperature inversions at $P < 0.5$ bar for WD-0137B \citep{Lothringer24}. Using this method, the day and night temperatures we calculated were $T_{\rm{Day}} = 2048.7\pm4.9$ and $T_{\rm{Night}} = 2027.9\pm4.8$.

From the grid search, the best-fitting combination of parameters resulted in $A_{\rm{B}} = 0.19^{+0.09}_{-0.03}$, $f_{\rm{irr-red}} = 0.81\pm0.08$, $T_{\rm{BD}} = 1810\pm70$ K, and $i=59.2^{+6.7}_{-1.3}$ degrees. In Figure~\ref{fig:tempmap}, we show the longitudinal temperature maps created with these values, in 3D and 2D projections.  The irradiation redistribution fraction value of $f_{\rm{irr-red}} = 0.81$ coupled with the small (19~K) and nearly constant day$-$night brightness temperature difference suggests an efficient day-to-night energy redistribution. This is further supported by the expected substellar temperature of $627$~K (assuming an $A_{\rm{B}} = 0.19$).

Another factor is the competition between radiative and circulatory timescales on the day-side. To an order of magnitude, the day-to-night circulatory timescale on GD\,1400B is half of the period: $t_{\rm{circ}}\sim\frac{1}{2}P\sim5$~hours. For the the radiative timescale, $\tau_{\rm{rad}}$, we assumed pressures between $P=1-50$ bars for the photosphere, typical for BDs at NIR wavelengths, and atmospheric temperatures between $T_{\rm{eff}}=2000-2200$~K. Estimating $\tau_{\rm{rad}}$ for a hydrogen atmosphere between these pressures and temperatures resulted in a range of $0.07 \gtrsim \tau_{\rm{rad}} \gtrsim 4.5$ hours\footnote{$\tau_{\rm{rad}} \sim \frac{P_{\gamma} c_p}{4 \sigma g T^3}$ where $P$ is the photospheric pressure, $c_p$ is the specific heat capacity of molecular H$_2$, $g$ is surface gravity, and $T$ is effective temperature. To estimate $\tau_{\rm{rad}}$ on the day-side of GD\,1400B, we used $P=1$ or $50$ bars, $c_p=13146.43$ m$^2$/s$^2$/K, log(g)=5.31, and $T=2000$ or $2200$ K.}. 

Comparing the timescales, radiative cooling on the day-side is slightly shorter than the circulation timescale for day-to-night energy redistribution. Consequently, a substantial fraction of the incident irradiation is likely emitted as radiation before it can be transported to the night-side. More detailed modeling of this atmosphere would provide valuable insights into the competition between timescales in this atmosphere, as well as the strength of day-to-night heat redistribution.

Perhaps most surprising is the non-irradiated BD temperature of $T_{\rm{BD}} = 1810\pm70$ K, which predicts the internal BD temperature in the absence of irradiation. Previous studies have estimated GD\,1400B to be spectral type $\sim$L\,$6 -$L\,$7$ \citep{Farihi05, Dobbie05}. Field L~dwarfs within this spectral type range typically have effective temperatures between $1441 - 1615$~K \citep{DupuyLiu17}, indicating that GD\,1400B is systematically too hot for its spectral type. 

However, at 0.074 M$_{\odot}$ ($\sim$77 M$_{\rm{Jup}}$), GD\,1400B is close to, and could well be above, the hydrogen burning-minimum mass as inferred from both observations \citep{DupuyLiu17} and theory \citep{Morley24_Diamondback}. This would explain the higher internal temperature predicted by the simple model and be in agreement with the measured day and night brightness temperatures around 2000~K. Whether a star or brown dwarf, at a total age of less than 2~Gyr \citep{Casewell24b}, GD\,1400B is still cooling from the time of its formation. \citet{Casewell24b} further explores this idea, proposing that GD\,1400B's cooling may still have been slowed by external irradiation.

\begin{figure}
    \begin{center}
    \includegraphics[width=0.48\textwidth]{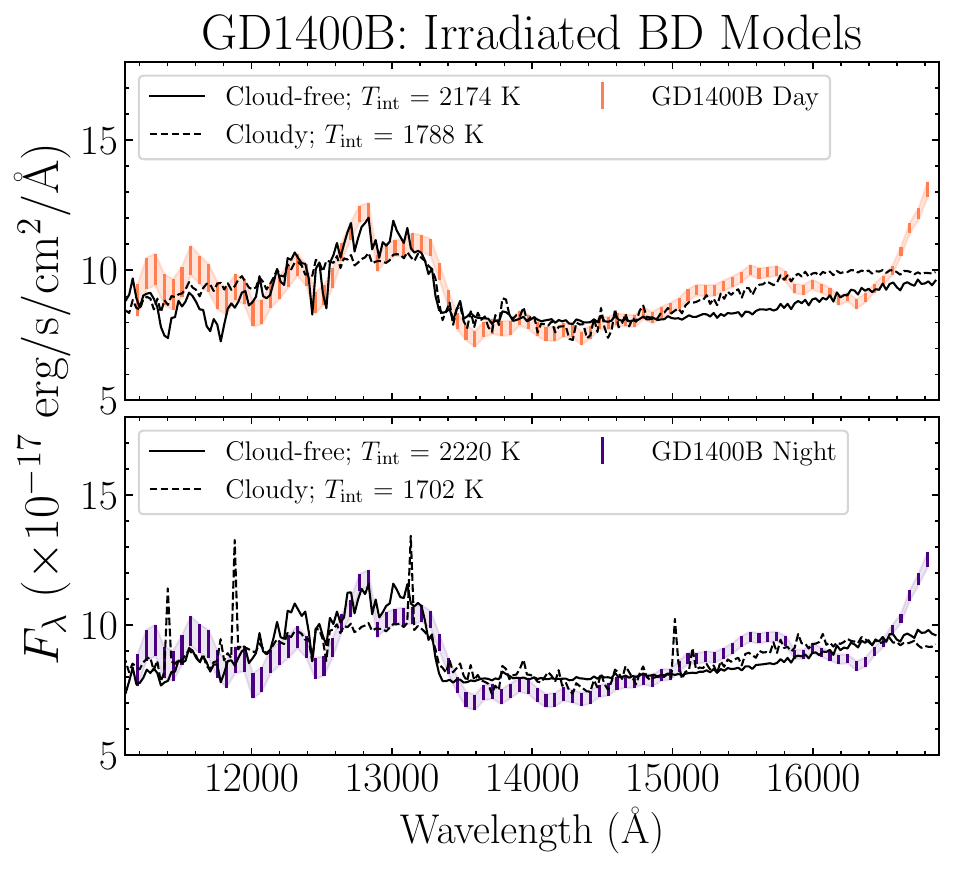}
    \caption{Comparison of the current best-fitting models from the PETRA retrievals (Section~\ref{sec:Compare_irr_models}) to the observations of GD\,1400B (orange and purple). Neither model fully captures all spectroscopic features, but a model with clouds (dashed line) provides a better fit than a cloud-free model (solid line) for both day and night hemispheres.}
    \label{fig:loth_models}
    \end{center}
\end{figure}

\begin{figure}
    \begin{center}
    \includegraphics[width=0.48\textwidth]{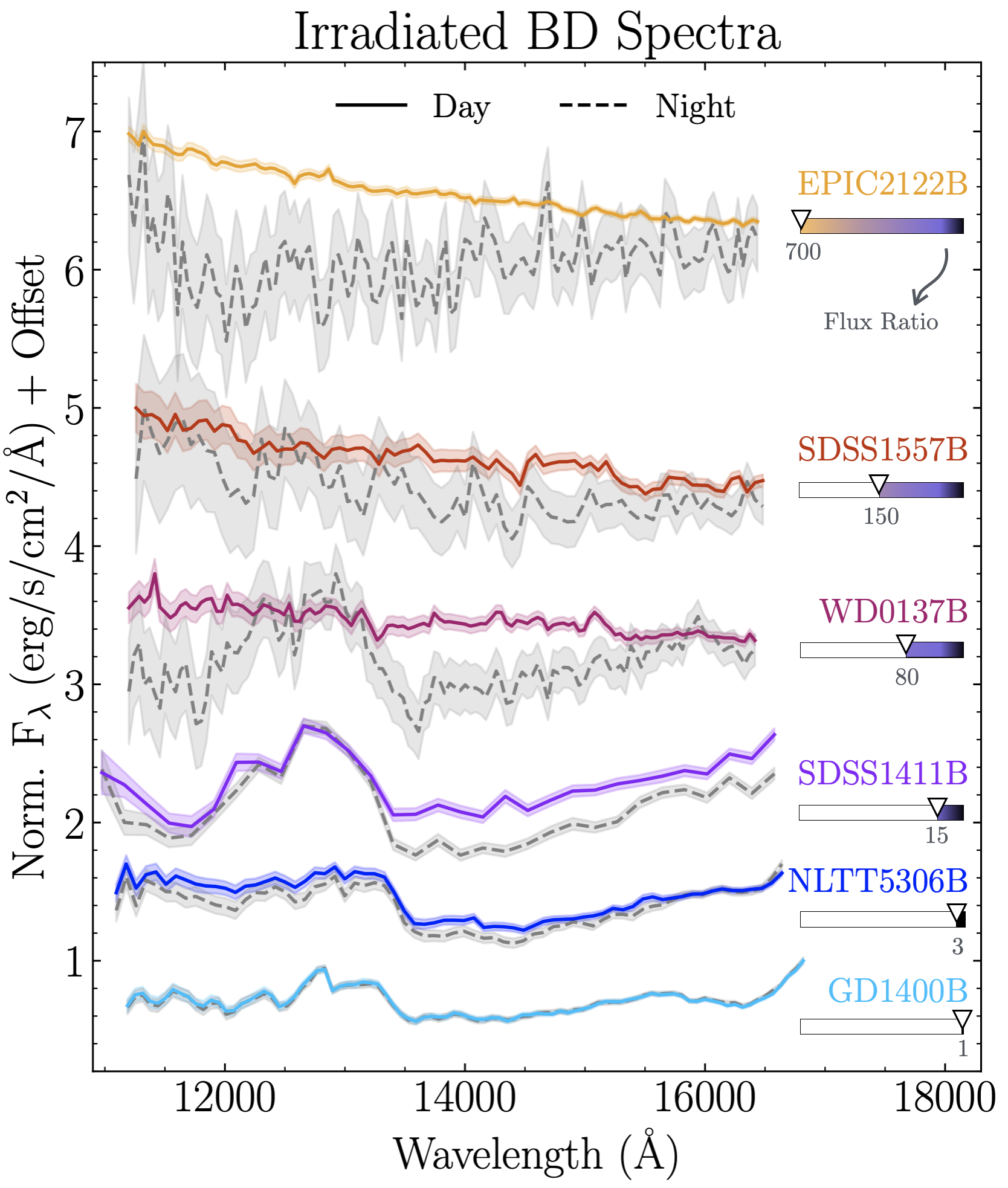}
    \caption{Day and night spectra for the Dancing with the Dwarfs (PI: Apai) WD--BD sample, ordered from most to least irradiated (top to bottom). Spectra were normalized to the median flux of each spectrum and offset for visual clarity. Each pair of spectra is labeled by the corresponding name on the right, with the relative flux ratio with respect to GD\,1400 depicted under each name. As irradiation flux decreases, we observed the day-side spectra transition from featureless to complex, with night-sides that appear to become more similar to the day-sides.}
    \label{fig:iBD_spectra}
    \end{center}
\end{figure}

\section{Comparison to Irradiated Brown Dwarf Models} \label{sec:Compare_irr_models}

Following retrievals of other irradiated brown dwarfs in \citet{Amaro24} and \citet{Lothringer24}, we fit GD\,1400's day and night side spectra using the PHOENIX Exoplanet Atmosphere Retrieval Algorithm \citep[PETRA,][]{LothringerBarman20_PETRA}, which uses the PHOENIX atmosphere model as the forward model in a Differential-Evolution Markov Chain Monte Carlo (DEMC) statistical framework. The temperature structure is parameterized using the 5-parameter profile from \citet{ParmentierGuillot14} with the addition of the internal temperature, $T_{\mathrm{int}}$ as a sixth free parameter. We also let the metallicity, [Fe/H], vary as a free parameter.

We initially ran retrievals without any cloud prescription, resulting in poor fits to the shape of the H$_2$O feature, as seen in Figure~\ref{fig:loth_models}. Additionally, [Fe/H] was retrieved to be super-solar at 0.9 $\pm$ 0.13 on the dayside and 0.9 $\pm$ 0.09 on the nightside. $T_{\mathrm{int}}$ is retrieved to be 2174~$^{+270}_{-89}$~K on the dayside and 2220~$^{+123}_{-45}$~K on the nightside.

Considering the temperatures in the atmosphere, we then ran retrievals with a grey cloud opacity, where the magnitude of the opacity and the cloud-top pressure were free parameters. As seen in Figure~\ref{fig:loth_models}, these retrievals fit the observed spectrum much better, with $\chi^2$ values improving from $>$100,000 for the cloud-free to $\sim$1000 for cloudy models. [Fe/H] for both the dayside and nightside are consistent with solar at 0.01 $\pm$ 0.38 and 0.24 $\pm$ 0.5. 

\begin{figure*}
    \begin{center}
    \includegraphics[width=0.98\textwidth]{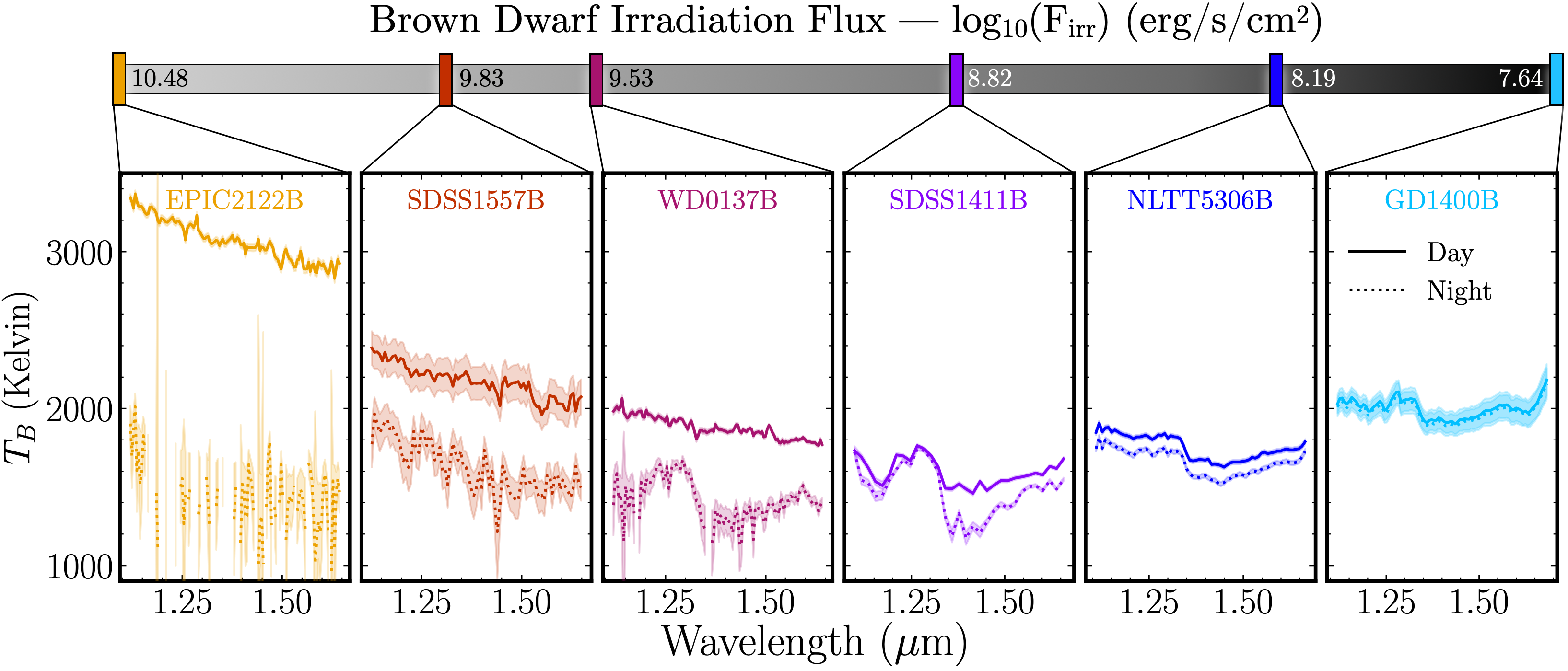}
    \caption{Brightness temperature sequence for the Dancing with the Dwarfs (PI: Apai) WD--BD sample, ordered from most to least irradiated (left to right). The total irradiation flux received by each BD is labeled on the bar above the brightness temperature panels, showing that our sample spans a significant range of irradiation fluxes.}
    \label{fig:iBD_brighttemps}
    \end{center}
\end{figure*}

Additionally, while the cloud opacity is not well-constrained, the cloud-top level is set between 50 mbar and 1 bar on the dayside and between 5 and 10 bars on the nightside. $T_{\mathrm{int}}$ is retrieved to be 1788~$^{+140}_{-216}$~K on the dayside and 1702~$^{+255}_{-168}$~K on the nightside.

These comparisons highlight the critical role of clouds in accurately modeling the spectra of GD\,1400. The significant improvement in fit quality when including a cloud prescription emphasizes the necessity of considering clouds for reproducing observed features. Additionally, the shift in retrieved metallicities from super-solar in the cloud-free models to values consistent with solar in the cloudy models suggests that neglecting clouds can lead to overestimations of atmospheric metallicity. Overall, these results demonstrate that clouds are very likely to be present in the atmosphere of GD\,1400.

\section{Full Irradiated Brown Dwarf Sample} \label{sec:dwd_sample}

GD\,1400 was the final WD--BD pair to be analyzed as part of the Dancing with the Dwarfs \textit{HST} observing program (PI: Apai). As such, we began investigating trends within the full irradiated BD sample, focusing on qualitative trends and leaving a more in-depth analysis for future work. Figure~\ref{fig:iBD_spectra} shows the normalized day and night spectra for each irradiated BD. For the three most irradiated BDs (EPIC2122B, SDSS1557B, and WD0137B), the day-side and night-side spectra present as nearly featureless slopes, whereas their less irradiated counterparts (SDSS1411B, NLTT5306B, and GD\,1400B), exhibit absorption features more reminiscent of isolated BDs, such as water absorption centered around 1.4 $\mu$m.

Additionally, the more irradiated pairs show the most dissimilarity between the features and higher flux contrasts between the day and night spectra. We observed that as external irradiation decreased, the night side spectra became more similar to the day sides. In fact, GD\,1400B, our least irradiated BD in the sequence, had nearly identical day and night spectra.

This prompted us to calculate the day and night brightness temperatures for each irradiated BD, shown in Figure~\ref{fig:iBD_brighttemps}. By arranging the brightness temperatures in order from the most (left) to the least (right) irradiated, we observed a pattern in the day-side temperatures, resembling a parabolic curve, with a noticeable decline from EPIC2122B to SDSS1411B and subsequent increase after SDSS1411B.

For the night-side brightness temperatures, there seemed to be a consistent range between 1200 and 2000 K for the four most irradiated BDs, with features becoming more prominent as irradiation decreased. However, after SDSS1411B, this trend changed, with night-side temperatures increasing to match the day-side temperatures and features once again becoming less prominent. This pattern might suggest that SDSS1411B experiences an external irradiation at or near a turn-around point for irradiated atmospheres.

\begin{figure*}
    \begin{center}
    \includegraphics[width=0.98\textwidth]{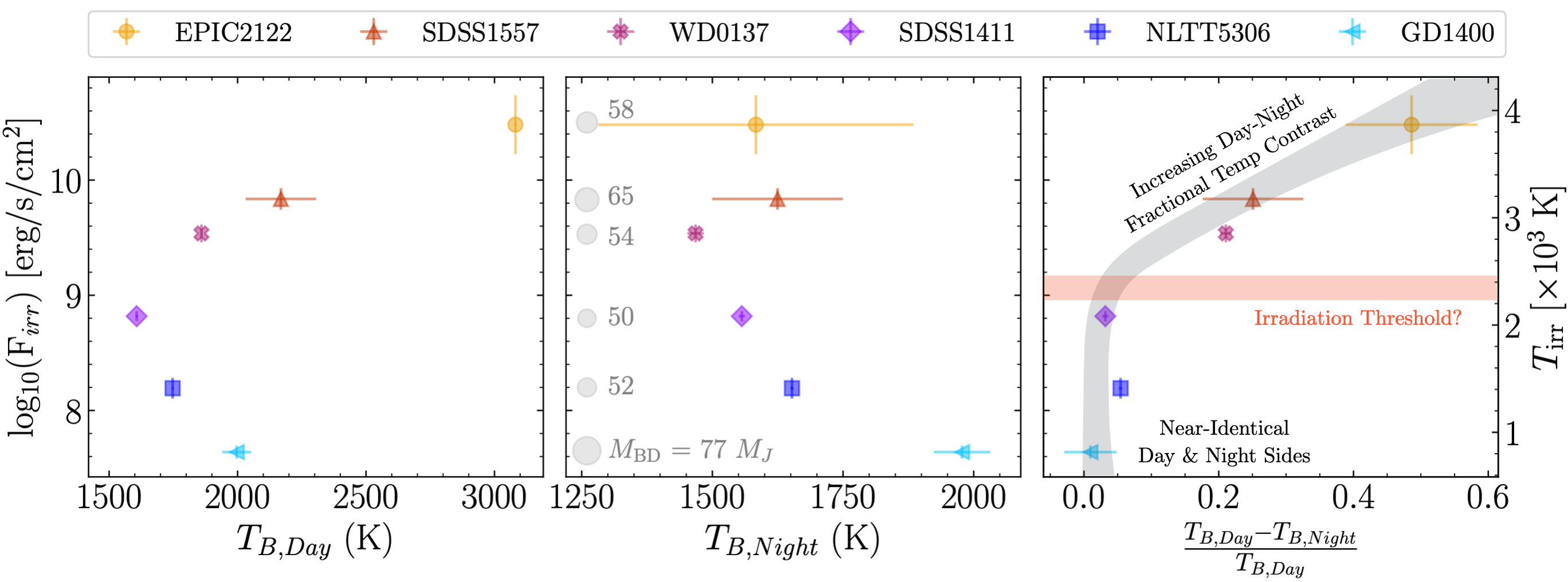}
    \caption{Comparison between irradiation strength and derived brightness temperatures for day (left) and night (middle) hemispheres as well as the fractional day$-$night contrast (right). With increasing irradiation, the day-side temperatures decrease from 2000 to 1600~K, up until 10$^9$ erg/s/cm$^2$. Above 10$^9$ erg/s/cm$^2$, day-side temperatures increase again. Night-side temperatures do not follow the same turn around. Instead, they generally decrease as irradiation strength increases, with the exceptions of EPIC2122B and SDSS1557B, the two most irradiated brown dwarfs in the sample. Brown dwarf masses are also labeled in this panel, to highlight that more massive brown dwarfs generally have hotter night-side temperatures, which is consistent with more massive brown dwarfs having hotter internal temperatures. Finally, the fractional day$-$night contrasts in the right-hand panel appear to be nearly zero until 10$^9$ erg/s/cm$^2$, when the contrasts increase again with higher irradiation. This may be explained by more efficient day-side radiative cooling at higher irradiation levels, which lowers the efficiency of day-to-night heat redistribution and increases the day-night temperature difference.}
    \label{fig:iBD_trends}
    \end{center}
\end{figure*}

To explore this possibility further, we conducted a sample-wide comparison between the day and night brightness temperatures, as well as fractional day$-$night temperature contrasts, and the external irradiation, shown in Figure~\ref{fig:iBD_trends}. The brightness temperatures used for this comparison are inverse variance-weighted averages across the G141 wavelength range. Interestingly, the day-side brightness temperatures appear to follow a quadratic trend, with a possible turn-around point at or near $10^9$ erg/s/cm$^2$.
This would follow a trend observed in hot Jupiters, where incident flux levels $>10^9$ erg/s/cm$^2$ coincide with cloud-free, gaseous atmospheres and thermal inversions \citep{Fortney08}. A quantitative comparison between the two populations would further validate irradiated brown dwarfs as suitable analogs for understanding hot Jupiter atmospheres.

The average night-side brightness temperatures did not follow a simple linear or quadratic trend; instead, most remained within a temperature range of 1400$-$1700~K, except for GD\,1400B, which had an observed night-side brightness temperature of approximately 2000~K. This trend may be correlated with the masses of the brown dwarfs, since higher mass BDs typically have higher internal temperatures. Comparing the published BD masses to the night-side temperatures in the middle panel of Figure~\ref{fig:iBD_trends}, a weak correlation may exist and would be beneficial to explore.

Conversely, the fractional day$-$night temperature contrasts (right panel of Figure~\ref{fig:iBD_trends}) exhibited two trends, once again above and below incident flux of 10$^9$ erg/s/cm$^2$. We find that above $F_{\rm{irr}}\approx10^9$, temperature contrasts increase with increasing irradiation, which is also observed in studies of hot Jupiters \citep{Perez-Becker13, Beatty19, Bell21}. This trend, greater day$-$night temperature contrasts may be explained by radiative cooling becoming stronger with increased irradiation on the day-side, which causes a decrease in day-to-night redistribution efficiency. Below $F_{\rm{irr}}\approx10^9$, temperature contrasts are nearly identical, and thus do not significantly vary with changing irradiation. This regime is consistent with efficient heat redistribution, i.e. uniform temperature day- and night-sides.

\section{Conclusions}
\label{sec:conclusions}

In this study, we presented time-resolved, spectrophotometric phase-mapping of the weakly irradiated brown dwarf GD\,1400B and initial comparisons between the full WD--BD \textit{HST} sample. The key findings of our study are as follows:

\begin{itemize}
    \item We presented high-quality \textit{HST}/WFC3/G141 phase-resolved spectra of white dwarf + brown dwarf binary GD\,1400. The observations sampled more than one full rotation of the BD and the WD shows signs of ZZ Ceti pulsations.
    
    \item We modeled the broad-band light curve with the lowest-order Fourier series combination, avoiding the higher-order ZZ Ceti pulsations. The best-fitting phase curve model yielded a relatively weak amplitude modulation of $\sim$1 percent.

    \item By dividing out the BD light curve model, we examined the ``in pulse'' spectra from each orbit and found no significant spectral differences (Figure~\ref{fig:pulse_spec}), consistent with ZZ Ceti pulsations in WDs. The largest ``pulse'' was nearly 4\%, which would result in a 0.74\% increase in irradiation temperature at the substellar point of the BD.
    
    \item We explored synthetic sub-band light curves in 2MASS $J$--, 1.4 $\mu$m Water--, and 2MASS $H'$--bands. We found that the $J$-- and Water--band amplitudes were similar (1.761$\pm$0.035 and 1.699$\pm$0.049\%) but the $H'$--band exhibited a larger amplitude (1.975$\pm$0.049\%). No significant wavelength-dependence on amplitude was observed.
    
    \item After modeling and subtracting the white dwarf contribution, we extracted day- and night-side spectra of GD\,1400B. We show that the two hemispheres are likely chemically similar, with a slightly bluer spectrum in the day-side. 
    
    \item The day and night hemisphere-averaged brightness temperatures show strong wavelength dependence, ranging between 1900 and 2100 K between 1.1 and 1.67 $\mu$m. The relatively higher temperatures and small differences between the hemispheres could either be due to highly efficient heat redistribution, an atmosphere where radiative escape strongly outcompetes atmospheric circulation, or both.

    \item A simple radiative and energy redistribution model closely reproduced the observed day and night brightness temperatures, predicting a $A_B$ = 0.19$^{+0.09}_{-0.03}$, $f_{\rm{irr-red}}$ = 0.81$\pm$0.08, $T_{\rm{non-irr}}$ = 1810$\pm$70 K, and $i$ = 59.2$^{+6.7}_{-1.3}$ for GD\,1400B.

    \item The day and night spectra of GD\,1400B were compared to one-dimensional models of irradiated brown dwarfs using PETRA \citep{LothringerBarman20_PETRA}. Models that included clouds were a better fit for the spectra, whereas cloudless models yielded poor fits to the water features and super-solar metallicities. Given the similar spectra and modeling results, we conclude that GD\,1400 likely has global cloud coverage. 

    \item With the Dancing with the Dwarfs (PI: Apai) observing program now complete, we began qualitatively exploring atmospheric trends across the six irradiated BDs. We revealed a possible irradiation threshold (10$^9$ erg/s/cm$^2$) for trends in absorption features and brightness temperatures. The trend we identify is consistent with efficient heat redistribution (isothermal day- and night-sides) for irradiation levels below the threshold, and decreasing efficiency, i.e. greater day-night temperature contrasts, above the threshold.

    \item A quantitative analysis of our spatially and spectrally resolved irradiated BD spectra, coupled with future JWST observations \citep{Zhou24_jwstprop_4967}, will help to answer the fundamental questions surrounding irradiated ultracool atmospheres.

\end{itemize}

\begin{acknowledgments}
This material is based upon work supported by the National Science Foundation Graduate Research Fellowship under Grant No. DGE-1746060. Any opinion, findings, and conclusions or recommendations expressed in this material are those of the authors(s) and do not necessarily reflect the views of the National Science Foundation. HST data presented in this paper were obtained from the Mikulski Archive for Space Telescopes (MAST) at the Space Telescope Science Institute. The specific observations analyzed can be accessed via \dataset[10.17909/rr9d-fj91]{https://doi.org/10.17909/rr9d-fj91}. Support for Program numbers HST-GO-15947 and HST-AR-16142 was provided by NASA through a grant from the Space Telescope Science Institute, which is operated by the Association of Universities for Research in Astronomy, Incorporated, under NASA contract NAS5-26555.
\end{acknowledgments}

\textit{Author contributions}: 
R.C.A. performed the data reduction, analysis, and drafted the manuscript.  D.A. developed the research project and made major contributions to the manuscript. Y.Z. and provided guidance and input regarding data reduction, light curve model fitting, and writing the manuscript. J.L. adapted existing 1D atmosphere models to our data and made contributions to the modeling sections of the manuscript. S.C., X.T., and VP provided guidance regarding the intellectual direction of analysis. All authors, including B.L., L.C.M., M.M., and T.B., contributed guidance and suggestions to improve the manuscript.
\vspace{5mm}
\facility{Hubble Space Telescope (WFC3)}

\software{astropy \citep{astropy2013,astropy2018, astropy2022},  
          Source Extractor \citep{Bertin96_sextractor},
          NumPy \citep{harris2020array},
          SciPy \citep{scipy2020-NMeth},
          Matplotlib \citep{Hunter07_matplotlib},
          HSTaXe (\url{https://github.com/spacetelescope/hstaxe})
          }



\bibliography{main}
\bibliographystyle{aasjournal}

\end{document}